\documentclass[a4paper,10pt,twocolumn]{article}

\usepackage[english]{babel}
\usepackage[utf8]{inputenc}
\usepackage[T1]{fontenc}

\usepackage[top=1.5cm, left=1.5cm, right=1.5cm, bottom=1.5cm]{geometry}

\renewenvironment{abstract}{\bf\small {\em\ Abstract---}}{}

\usepackage{amsfonts,amssymb,amsmath,amsthm}
\usepackage{subfigure}
\usepackage{graphicx}
\usepackage[footnotesize]{caption}

\usepackage{amsfonts,amssymb,amsmath,amsthm}

\title{A Low-Rank and Joint-Sparse Model for Ultrasound \\ Signal Reconstruction}

\author{Miaomiao Zhang$^1$, Ivan Markovsky$^2$, Colas Schretter$^2$ and Jan D'hooge$^{1}$\\
  \footnotesize $^1$Cardiovascular Sciences, KU Leuven, Leuven, Belgium.\ $^2$ Vrije Universiteit Brussel (VUB), Brussels, Belgium.  } \date{\empty} 

\begin{document}

\maketitle

\begin{abstract} 
	With the introduction of very dense sensor arrays in ultrasound (US) imaging, data transfer rate and data storage became a bottleneck in ultrasound system design. To reduce the amount of sampled channel data, we propose to use a low-rank and joint-sparse model to represent US signals and exploit the correlations between adjacent receiving channels.
	Results show that the proposed method is adapted to the ultrasound signals and can recover high quality image approximations from as low as 10\% of the samples.

\end{abstract}

\section{Introduction}
\label{sec:introduction}
Ultrasound echography is one of the most used diagnostic imaging techniques as it is real-time, safe, low-cost and portable. Conventional ultrasound imaging is usually performed by scanning a medium using sequential focused beams, each firing allowing the reconstruction of one line of the final image, \textit{i.e.} Single-Line-Transmission (SLT) imaging. A high-resolution image requires sufficient numbers of transmissions as well as a sampling rate that is significantly higher than the Nyquist rate of the signal~\cite{DigitalBF}. Consequently, with such high sampling rate, and taking
into account the number of transmissions and the number of
transducer elements, the amount of sampled data can become
enormous, which restricts the data storage and transportation
in most of the commercial systems today. In this context, several strategies based on compressed sensing (CS) have been proposed \cite{Colas}. However, the reconstruction accuracy of CS-based methods is highly dependent on the basis and the number of measurements for each channel cannot be lower than the sparsity thereby limiting the data reduction rate. The aim of the current study was therefore to further reduce data rates by exploiting the low-rank property of US signals.

\section{Methodology}
\label{sec:first-section}

Let us rearrange all the received pre-beamformed data in a 2D matrix $\textbf{X} \in\,\mathbb{R}^{M\times\,N}$. According to the fact that the pre-beamformed US signals from different transducer elements are joint-sparse in the Fourier domain ~\cite{DCS}, $\textbf{X}$ can be expressed in a matrix form as:
\begin{equation}
\label{eq:Fourier_model_1D_2}
\textbf{X} = \textbf{YD}
\end{equation}
where $\textbf{Y} \in\,\mathbb{R}^{M\times\,k}$ is a partial 1D Fourier matrix with $k (k\ll M)$ frequencies inside the bandwidth. This hypothesis is reasonable,
since the US radio frequency (RF) signals are bandlimited by the
impulse response of the transducer. Assuming that the signal bandwidth is 1, the $k = M(2f_c/f_s)$ frequencies are the frequencies inside $[f_c/2,3f_c/2]$ and $[-f_c/2,-3f_c/2]$, where $f_c$ and $f_s$ are the center and sampling frequency of the RF signal. Thus $\textbf{D} \in\,\mathbb{R}^{k\times\,N}$ is the corresponding 1D Fourier coefficient matrix of the signal at those $k$ frequencies. In addition, since the maximum bandwidth of the RF signal is 1, the real joint-sparsity $K_k$ of the signal in Fourier domain is no more than $k$, \textit{i.e.} $K_k \leq k$.  

The factorization form in (\ref{eq:Fourier_model_1D_2}) implies a low-rank structure of $\textbf{X}$ with:
\begin {equation}
\text{rank}(\textbf{X})\leq\,k\,\,\space\space\,  \text{when}\space\,\, k<N 
\end{equation}
where $ \text{rank}(\textbf{X}) = k$ if and only if $\textbf{D}$ is full-rank. In practice, thanks to the strong correlation between US signals, we have $ rank(\textbf{X}) \ll k$. Thus, $\textbf{X}$ has a low-rank and joint-sparse structure when we have enough number of US channel signals, \textit{i.e.} $N > \text{rank}(\textbf{X})$. Fortunately, in the field of US, the above condition is generally satisfied with $N\gg\,\text{rank}(\textbf{X})$, implying that the low-rank and joint-sparse property can be applied in US signal reconstruction.

Assuming that the data is contaminated by random noise, the measurements $\textbf{B}$ can be modeled as :
\begin{equation}
\label{eq:samples_2D}
\textbf{B} = \textbf{P}_\Omega(\textbf{X})+\textbf{N}_e
\end{equation}
where $\Omega$ is a set of locations where the signal $\textbf{X}(m,n)$ is observed, \textit{i.e.} $\textbf{X}(m,n)$ known if $(m,n) \in\Omega$. $\textbf{P}_\Omega(\textbf{X})$ represents the corresponding values of $\textbf{X}$ in the locations of $\Omega$ and $\textbf{N}_e$ is the additive noise term. In these settings, the reconstruction problem thus amounts to solve (\ref{eq:samples_2D}) for $\textbf{X}$, under the constraint that $\textbf{X}$ is low-rank and joint-sparse. This problem could be reformulated as an unconstrained optimization problem as in ~\cite{Golbabaee}: 
\begin{equation}
\label{eq:problem_1}
\hat{\textbf{X}} = arg\min_{\textbf{X}} \left|\left|\textbf{X}\right|\right|_{\ast}+\alpha\left|\left|\textbf{Y}_t\textbf{X}\right|\right|_{2,1}+\frac{1}{2\mu}\left|\left|\textbf{B}-\textbf{P}_\Omega(\textbf{X})\right|\right|_F^2
\end{equation}
where $\left|\left|\textbf{X}\right|\right|_\ast = \sum_i\sigma_i$ is the sum of the singular values (\textit{i.e.} the nuclear norm) that aims at imposing the low-rank property of the RF signal $\textbf{X}$. This assumption has been extensively used in matrix completion, see \cite{MC,Ivan}; $\textbf{Y}_t$ is the adjoint operator of $\textbf{Y}$ with the relation $\textbf{Y}_t\textbf{X} = \textbf{D} $. In our case, $\textbf{Y}_t$ and $\textbf{Y}$ are the Fourier and Inverse Fourier matrix with effective frequencies $f_i$; $\left|\left|\textbf{D}\right|\right|_{2,1} = \sum_{q=1}^{k}\left|\left|\textbf{D}^{q\rightarrow}\right|\right|_2 $ (\textit{i.e.} the $l_{2,1}$ norm) that is used to explore the joint sparsity property of $\textbf{D}$  and $q\rightarrow$ denotes the $q$-th row; $\left|\left|\cdot\right|\right|_F $ is the Frobenius norm. The parameters $\alpha$ and $\mu$ give the trade-off among the nuclear norm term $\left|\left|\textbf{X}\right|\right|_\ast$, the $l_{2,1}$ norm term $\left|\left|\textbf{Y}_t\textbf{X}\right|\right|_{2,1}$ and the data consistency term $\left|\left|\textbf{B}-\textbf{P}_\Omega(\textbf{X})\right|\right|_F^2$. Fundamentally, we are looking for a matrix with minimum rank and joint-sparsity subject to the acquired data. Since $\textbf{X}$ has the same rank as $\textbf{D}$, this model could be reformulated as: 
\begin{equation}
\label{eq:problem_2}
\hat{\textbf{D}} = arg\min_{\textbf{D}} \left|\left|\textbf{D}\right|\right|_{\ast}+\alpha\left|\left|\textbf{D}\right|\right|_{2,1}+\frac{1}{2\mu}\left|\left|\textbf{B}-\textbf{P}_\Omega(\textbf{YD})\right|\right|_F^2
\end{equation}

\begin{figure*}[hbt]
	\centering
	\includegraphics[width=16cm]{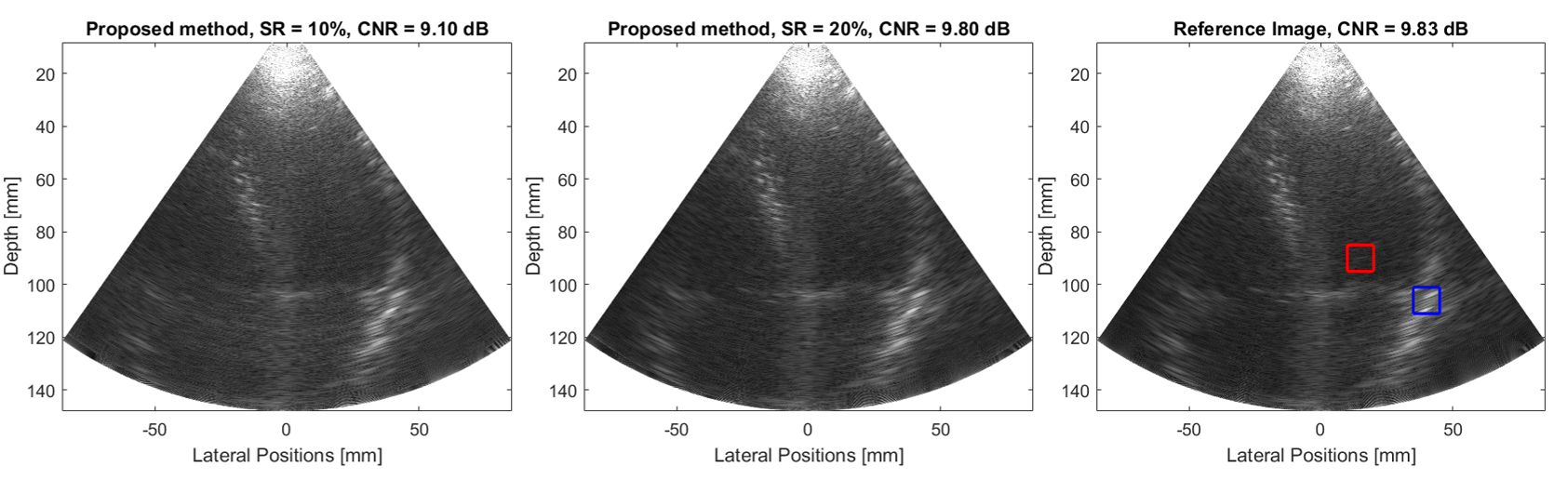}
	\caption{Original \textit{in vivo} cardiac image and the images reconstructed by the proposed method with different sampling rates. The red and blue block in the reference image represent the regions used to compute CNR.}
	\label{fig:simu_images}
\end{figure*}

The objective function in (\ref{eq:problem_2}) aims at estimating $\textbf{D}$ instead of $\textbf{X}$ directly from the acquired samples. It is worth to notice that $\textbf{D}$ is a $k\times\,N$ matrix with $k\ll M$, which means the number of variables to be estimated in $\textbf{D}$ is $\frac{M}{k}$ times less than $\textbf{X}$ and the problem (\ref{eq:problem_1}) is much simplified. To solve the optimization problem in (\ref{eq:problem_2}), we proposed in the following section an algorithm based on simultaneous direction method of multipliers (SDMM) \cite{SDMM}.

\section{Algorithm}
In this section, we adapted the SDMM optimization framework to solve the problem in (\ref{eq:problem_2}), which iteratively solves the above optimization problem as follows:

\textbf{Step 1:} Updating $\textbf{D}$, given $\textbf{b}_i$ and $\textbf{w}_i$:
\begin{equation}
\label{eq:sub_1}
\textbf{D}^{s+1} = \arg\min_{\textbf{D}}\frac{1}{2\gamma}\left|\left|\begin{pmatrix}
\textbf{b}_1^s \\
\textbf{b}_2^s\\
\textbf{b}_3^s \\   
\end{pmatrix}+\begin{pmatrix}
\textbf{I} \\
\textbf{I}\\
\textbf{Y}\\   
\end{pmatrix}\textbf{D}-\begin{pmatrix}
\textbf{w}_1^s \\
\textbf{w}_2^s\\
\textbf{w}_3^s \\   
\end{pmatrix}\right|\right|_F^2
\end{equation}
where $\textbf{w}_1 = \textbf{w}_2 = \textbf{D}$ and $\textbf{w}_3 = \textbf{YD}$ are setting to indicate the three convex objective functions, $\textbf{b}_1$, $\textbf{b}_2$ and $\textbf{b}_3$ are Lagrangian parameters that have the same dimensions as $\textbf{w}_1$, $\textbf{w}_2$ and $\textbf{w}_3$, respectively. $(\cdot)^s$ means the updated value of $(\cdot)$ from the $s$-th iteration. $\gamma>0$ is a penalty parameter.
As stated in \cite{SDMM}, (\ref{eq:sub_1}) is a classical $l_2$ norm minimization problem and can be efficiently solved.

\textbf{Step 2:} Solving $\textbf{w}_i$ using $\textbf{b}_i$ and $\textbf{D}$:
\begin{multline}
\label{eq:sub_2}
\begin{pmatrix}
\textbf{w}_1^{s+1} \\
\textbf{w}_2^{s+1}\\
\textbf{w}_3^{s+1} \\   
\end{pmatrix}= \arg\min_{\textbf{w}_1,\textbf{w}_2,\textbf{w}_3}\left\{\frac{1}{2\gamma}\left|\left|\begin{pmatrix}
\textbf{b}_1^s \\
\textbf{b}_2^s\\
\textbf{b}_3^s \\   
\end{pmatrix}+\begin{pmatrix}
\textbf{I} \\
\textbf{I}\\
\textbf{Y}\\   
\end{pmatrix}\textbf{D}^{s+1} \right.\right.\right.\\
\left.\left.\left.-\begin{pmatrix}
\textbf{w}_1 \\
\textbf{w}_2\\
\textbf{w}_3 \\   
\end{pmatrix}\right|\right|_F^2
+\sum_{i=1}^{3}g_i(\textbf{w}_i)\right\}
\end{multline}
where $ g_1(\textbf{w}_1) = \left|\left|\textbf{w}_1\right|\right|_\ast $, 
$ g_2(\textbf{w}_2) =  \alpha\left|\left|\textbf{w}_2\right|\right|_{2,1} $ and
$ g_3(\textbf{w}_3) =  \frac{1}{2\mu}\left|\left|\textbf{B}-\textbf{P}_\Omega(\textbf{w}_3)\right|\right|_F^2 $ are the three convex terms of the optimization problem. Due to the separate structure of (\ref{eq:sub_2}), it can be solved by minimizing the three subproblems that corresponding to the update of $\textbf{w}_1$, $\textbf{w}_2$ and $\textbf{w}_3$, respectively.

\textbf{Step 3:} Updating $\textbf{b}_i$, given $\textbf{D}$ and $\textbf{w}_i$:

\begin{equation}
\label{eq:sub_3}
\begin{pmatrix}
\textbf{b}_1^{s+1} \\
\textbf{b}_2^{s+1}\\
\textbf{b}_3^{s+1} \\   
\end{pmatrix} = \begin{pmatrix}
\textbf{b}_1^s \\
\textbf{b}_2^s\\
\textbf{b}_3^s \\   
\end{pmatrix}+\begin{pmatrix}
\textbf{I} \\
\textbf{I}\\
\textbf{Y}\\   
\end{pmatrix}\textbf{D}^{s+1}-\begin{pmatrix}
\textbf{w}_1^{s+1} \\
\textbf{w}_2^{s+1}\\
\textbf{w}_3^{s+1} \\   
\end{pmatrix}
\end{equation}


\section{Results}
\label{sec:second-section}

\begin{figure}[h]
	\centering
	\includegraphics[width=5cm]{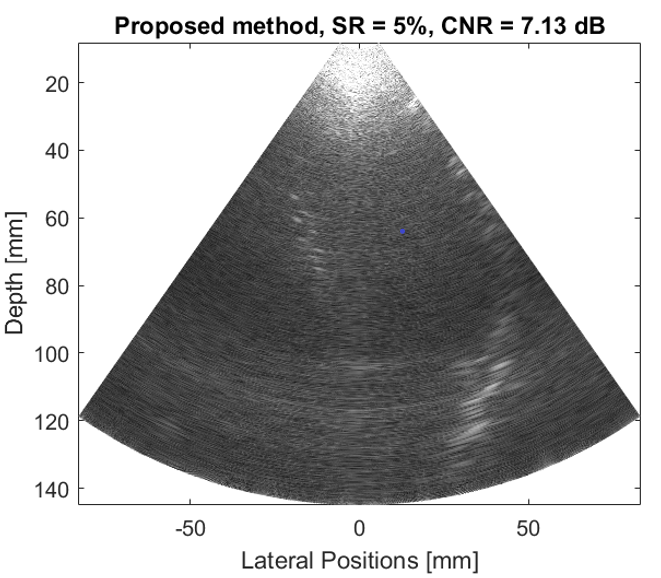}
	\caption{Image reconstructed by the proposed method with 5\% of samples.}
	\label{fig:simu_images_0.5}
\end{figure}

To validate the proposed methodology, \textit{in vivo} cardiac pre-beamformed RF data ($f_c$-3.5 MHz; $f_s$-25 MHz, matrix size $4838 \times 4544$) was captured using an experimental ultrasound system and reconstructed from a subset of samples obtained at different sampling rates (SR). 
The parameters in the optimization
algorithm were chosen as $\{\gamma, \alpha, \mu\} = \{1, 0.01, 1e-6\}$ by cross
validation. The contrast-to-noise-ratio (CNR) was calculated on the 8-bit grayscale B-mode image.

 Figure 1 and Figure 2 show the B-mode images reconstructed for different sampling rates together with the corresponding CNR. Strong artifacts appear for the reconstructed image with a sampling rate of 5\%, corresponding to a reduction in CNR of 2.7 dB. However, when the sampling rate increases, the
artifacts disappear and the CNR improves, resulting in images
visually very similar to the reference image. Image quality remained good even with only 10\% of the samples. In comparison, given the sparsity of the problem (\textit{i.e.} $2f_c/f_s$), using solely the sparsity promoting prior maximally achieve a SR of 28\%. 

\section{Conclusion}
\label{sec:conclusion}

This paper introduces a new data model based on the low-rank and joint-sparse priors. This regularization strategy allows for estimating missing data and thus reducing drastically the sampling rate. 
\textit{In vivo} experiment was performed to validate and evaluate the proposed method. The results demonstrate that the proposed approach is capable
of reconstructing the whole image from a sparse set of samples (\textit{e.g.} 10\% of samples) while keeping adequate image quality. However, the reconstruction algorithm is time consuming, thus real-time imaging is not realistic. Fast algorithm will be devised in future work.


\end{document}